\documentclass{kapproc} 
\let\footnote\savefootnote
\let\footnotetext\savefootnotetext 
\let\footnoterule\savefootnoterule 
\kluwerbib

\RequirePackage{graphicx}%
\RequirePackage{epsf}%

\begin{document}

\articletitle{Effects of Noise on Galaxy Isophotes}

\author{Nurur Rahman and Sergei F.\ Shandarin}

\affil{Department of Physics and Astronomy, University of Kansas, \\
Lawrence, KS 66045, USA}
\email{nurur@kusmos.phsx.ukans.edu, sergei@ukans.edu}

\begin{abstract}
The study of shapes of the images of objects is an important issue 
not only because it reveals its dynamical state but also it helps 
to understand the object's evolutionary history. 
We discuss a new technique in cosmological image analysis which 
is based on a set of non-parametric shape descriptors known as the 
Minkowski Functionals (MFs). These functionals are extremely versatile 
and under some conditions give a complete description of the geometrical 
properties of objects. We believe that MFs could be a useful tool to 
extract information about the shapes of galaxies, clusters of galaxies 
and superclusters. The information revealed by MFs can be 
utilized along with the knowledge obtained from currently popular 
methods and thus could improve our understanding of the true shapes 
of cosmological objects.
\end{abstract}
\section{Introduction}
The shape of an object provides important clues for understanding its 
nature, in particular, the past and  ongoing physical processes responsible 
for shaping the object. 
A mathematical branch known as the integral geometry offers an unique 
opportunity providing a set of simple morphological measures that can be 
used to characterize an isolated single objects as well as a multi-component 
object. 
These measures were suggested by H. Minkowski (Minkowski 1903) and known as 
the Minkowski Functionals (MFs). 
The parameters devised from MFs are robust, yielding local as well as global 
morphological information of any spatial structure.
We discuss this set of measures in galaxy morphology specifically we propose 
a technique to restore galaxy images distorted by the background noise. 
The MFs was introduced into cosmology by Mecke, Buchert, \& Wagner (1994) 
and later used in variety of cosmological problems (see e.g. Beisbart 2000; 
Beisbart, Buchert, \& Wagner 2001; Novikov, Feldman \& Shandarin, 1999; 
Schmalzing \& Buchert 1997; Schmalzing \& Gorski 1998; 
Schmalzing et al. 1999; Shandarin 2002; Shandarin et al. 2002; 
Shani, Sathyaprakash \& Shandarin 1998). 
\section{Reduction of Noise Distortion}
Background noise distorts the original profile of a galaxy and therefore 
its isophotal contours will always be deviated from their true shapes. 
In galaxy morphology finding the true shape of a galaxy profile is extremely 
important to understand its evolution, dynamical state and environmental 
effects.   
The effects of noise can be reduced in several ways. 
One approach is to smoothing the map with some filter, the other could 
be incorporating systematic effects into estimates of the parameters by 
introducing corrections.
We introduce a simple linear technique that smoothes not the whole map but 
only its contours chosen some specific levels above background. 
The method of smoothing is based on replacing the set of contour points by a 
new set each point of which is placed exactly in the middle of two adjacent 
points in the original set. This set of new points construct a new contour 
what we call a contour one time smoothed from the previous one. The Minkowski 
parameters are then computed for this smoothed contour. The procedure is 
applied iteratively many times depending on the length of the contour and 
the level of the noise. 
To implement this method, finding a lower limit to begin smoothing and an 
upper limit up to which one should stop is very crucial as without any 
knowledge of the true shape one can in fact smooth a contour so much that 
for using this particular approach the resulting contour would ultimately 
appear as circular and eventually with further smoothing as a point. To 
get a control over this situation one can convolve simulated galaxy profiles 
with noises of known properies, find effective limits for smoothing and 
then calibrate it for real galaxy images.
We follow this approach to select smoothing limits where we clip the galaxy 
image at a certain level above the background noise and find the 
total number of contour points, $N$. Next we set the first smoothing limit 
equal to the one-tenth ($N/10$) of the number of contour points. Detail 
analysis shows that this is a reasonable choice to start the smoothing. 
The subsequent higher number of smoothing is just the integer multiple of 
the first. 
Note that the required smoothing and the accuracy with which one can 
determine the original shape depends crucially on the steepness of the 
profile. Higher the gradient of the profile more accurate one would be to 
find the true shape and less amount smoothing will be required.

\begin{figure}
\epsscale{0.50}
\plotone{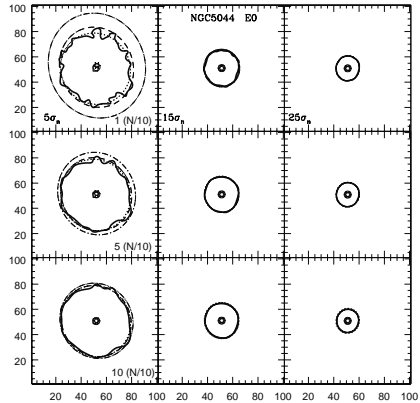}
\caption{A representative example of a galaxy's smoothed isophotal contour 
shown by the heavy solid line.
Three different points (open) circle, square and star represent three
vector MFs (a measure of asymmetry) and the three closed contours shown 
by the dotted, dashed, and dashed-dot lines are constructed from three 
tensor MFs (see Beisbart 2000).  
The 2MASS E0 galaxy NGC5044 at different rms noise levels (5, 15, 
25$\sigma_n$, from left to right) after certain amount of smoothing 
applied to contour points shown from top to bottom. 
The wiggling due the background noise is reduced by the contour smoothing 
as shown in the bottom panel. It also removes any apparent asymmetry present 
and makes the tensor ellipses to converge with one another. 
}
\end{figure}


\begin{chapthebibliography}{1}
\bibitem{Beisbart 2000}
Beisbart, C., 2000, Ph.D. Thesis, Ludwig-Maximilians-Universit\"{a}t, 
Munich  

\bibitem{Beisbart, Buchert \& Wagner 2001}
Beisbart, C., Buchert, T., \& Wagner, H., 2001a, Physica A, 293, 529B

\bibitem{Mecke, Buchert, \& Wagner 1994}
Mecke, K. R., Buchert, T., \& Wagner, H., 1994, A\&A, 288, 697

\bibitem{Minkowski 1903}
Minkowski, H., 1903, Math. Ann., 57, 447

\bibitem{Novikov, Feldman \& Shandarin 1999}
Novikov, D., Feldman, H. A., \& Shandarin, S. F., 1999, 
Int. J. Mod. Phys., D8, 291

\bibitem{Schmalzing \& Buchert 1997}
Schmalzing, J., \& Buchert, T., 1997, ApJ, 482, L1

\bibitem{Schmalzing \& Gorski 1998}
Schmalzing, J., \& Gorski, K. M., 1998, MNRAS, 297, 355

\bibitem{Schmalzing et al. 1999}
Schmalzing, J., Buchert, T., Melott, A. L., Shani, V., 
Sathyaprakash, B. S., \& Shandarin, S. F., ApJ, 526, 568

\bibitem{Shandarin 2002}
Shandarin, S.F., 2002, MNRAS, 331, 865

\bibitem{Shandarin et al. 2002}
Shandarin, S.F., Feldman, H. A., Xu, Y., \& Tegmark, M., 2002, 
ApJS, 141, (astro-ph/0107136)

\bibitem{Shani, Sathyaprakash \& Shandarin 1998}
Shani, V., Sathyaprakash, B. S., \& Shandarin, S. F., 1998, 
ApJ, 495, L5
\end{chapthebibliography}
\end{document}